\documentclass[twocolumn,nofootinbib,amsfonts,prd,aps]{revtex4}

\usepackage{amsmath,amssymb,amsfonts,amsthm,mathrsfs}
\usepackage[pdftex]{hyperref}
\usepackage{color}

\def\be{\begin{equation}}
\def\ee{\end{equation}}
\def\ba{\begin{eqnarray}}
\def\ea{\end{eqnarray}}

\def\qh{\hat{q}}
\def\xh{\hat{x}}
\def\w{\omega}
\def\zb{\bar{z}}

\def\Qs{Q^{\rm soft}}
\def\Qh{Q^{\rm hard}}
\def\Q{\mathcal{Q}}
\def\Qcs{\Q^{\rm soft}}
\def\Qch{\Q^{\rm hard}}
\def\tf{\text{TF}}

\begin{document}
\title{Sub-subleading soft gravitons: New symmetries of quantum gravity?}

\author{Miguel Campiglia$^{1}$, Alok Laddha$^{2}$}
\affiliation {
1. Instituto de F\'{\i}sica, Facultad de Ciencias, 
Igu\'a 4225,  11400 Montevideo, Uruguay. \\
2. Chennai Mathematical Institute, Siruseri 603103, India}

\begin{abstract}
Due to seminal works of Weinberg, Cachazo and Strominger we  know that tree level quantum gravity amplitudes satisfy three factorization constraints. Building on  previous works which relate two of these  constraints to symmetries of quantum gravity at null infinity, we present rather strong evidence  that the third constraint is also equivalent to a new set of symmetries of (perturbative) quantum gravity. Our analysis implies that the  symmetry group of quantum gravity may be even richer than the BMS group (or infinite dimensional extension thereof) previously considered.
\end{abstract}

\maketitle

There has been  significant recent progress in our understanding of symmetries associated to
quantum gravity in asymptotically flat spacetimes. We now understand that at least at perturbative level, these symmetries contain an infinite dimensional group which is one of  two possible extensions of the Bondi-Metzner-Sachs (BMS) group \cite{bms}  that has long been known to be a symmetry of classical general relativity.\footnote{See \cite{aa} for the fundamentals of BMS in quantum gravity.}  The key evidence for these
groups come from their relation with certain soft theorems in (perturbative) quantum gravity. In particular
as shown in \cite{stromST}, the statement that supertranslations are symmetries of the quantum 
gravity S matrix is encoded in Weinberg's soft graviton theorem \cite{weinberg}. 
 In \cite{stromvir} this idea was extended to the Cachazo-Strominger subleading soft theorem \cite{cs} where it was argued it implied a Virasoro symmetry of  locally conformal Killing vector fields of the sphere at null infinity \cite{barnich}. Based on these works we showed  \cite{cl1} that the subleading soft theorem can alternatively   be understood as the statement that  the group of  diffeomorphisms of the sphere at null infinity is a symmetry of the gravitational S matrix. 

The soft theorems in themselves are rather fascinating statements. As argued in \cite{cs},  when in a gravitational scattering
process one of the gravitons becomes `soft' (its energy goes to zero), 
the tree level
scattering amplitude factorizes upto first order 
in the soft 
graviton energy $E_q$:
\begin{multline}
{\cal M}_{n+1}(k_{1},\dots,k_{n};q) = \\ (E^{-1}_q S^{(0)} + S^{(1)} + E_q S^{(2)}){\cal M}_{n}(k_{1},\dots,k_{n})  +O(E_q^2). \label{softexpansion}
\end{multline}
Since both $S^{(0)}$ and $S^{(1)}$ are 
associated to (one of the two possible) extensions of BMS, 
it is a natural question to ask if the final  factorization term, namely $S^{(2)}$, 
is also associated to 
symmetries of  quantum gravity.  In this paper we present rather strong evidence 
that this is the case. As summarized below, our strategy involves looking at the known results from a slightly new perspective \cite{maxwell} that includes a dual or `magnetic' version of the usual charges. 

The expansion (\ref{softexpansion}) yields the three equations:
\ba
&\displaystyle\lim_{E_q \rightarrow 0} E_q {\cal M}_{n+1}  &=  S^{(0)}{\cal M}_{n} \label{S0id}\\
 &\displaystyle\lim_{E_q \rightarrow 0}  {\cal M}_{n+1} |_{\textrm{fin}} & =  S^{(1)}{\cal M}_{n} \label{S1id}\\
 &\displaystyle\lim_{E_q \rightarrow 0} E^{-1}_q {\cal M}_{n+1} |_{\textrm{fin}} &=  S^{(2)}{\cal M}_{n} \label{S2id}
\ea
(in the last two lines one keeps the finite piece and discard terms proportional to  $E^{-1}_q$ and $E^{-2}_q$). Since the emitted soft graviton has two possible polarizations, each of these equations provides two independent identities (per point on the sphere of soft graviton directions). One would like  to realize such identities as Ward identities associated to appropriate charges. In \cite{stromST} it was shown that (\ref{S0id}) corresponds to supertranslations Ward identites:
\be
\langle\textrm{out}\vert [Q_{f},S] \vert\textrm{in}\rangle\ =\ 0, \label{stwardid}
\ee
where $Q_{f}$ is the charge associated to a supertranslation vector field $\xi^a \sim f \partial_u$ and   $\langle\textrm{out}\vert S \vert\textrm{in}\rangle\ = {\cal M}_{n}(k_{1},\dots,k_{n})$. Now, since  (\ref{stwardid}) is parametrized by  functions on the sphere $f$, it counts as one identity per point on the sphere. Where is the second identity? In \cite{stromST} this second identity is associated to certain Christodoulou-Klainerman (CK) condition imposed on the free data \cite{strom1}. Now, it turns out  that this second condition may also be realized as  Ward identities of `dual' supertranslation charges \cite{long}:
\be
\langle\textrm{out}\vert [Q^*_{f},S] \vert\textrm{in}\rangle\ =\ 0. \label{magstwardid}
\ee
Here $Q^*_{f}$ is the `magnetic' version of $Q_f$ that is obtained by dualizing the Weyl tensor \cite{nut4,long}.  Thus, the two identities contained in (\ref{S0id}) are equivalent to the two identities (\ref{stwardid}) and (\ref{magstwardid}).

In \cite{cl1} we showed that (\ref{S1id}) is equivalent to certain $\textrm{Diff}(S^2)$ Ward identities,
\be
\langle\textrm{out}\vert [Q_{V},S] \vert\textrm{in}\rangle\ =\ 0. \label{vfwardid}
\ee
associated to `generalized BMS'  vector fields $\xi^a \sim V^A \partial_A$. In this case the charges are parametrized by arbitrary sphere vector fields $V^A$ and  (\ref{vfwardid}) corresponds to two identities per sphere point. This was a key point in showing the equivalence with (\ref{S1id}) without further CK-type conditions. What about the magnetic version of (\ref{vfwardid})? It turns out \cite{long} that in this case $Q^*_{V^A}= Q_{{\epsilon}^{A}_B V^B}$ and hence no further charges arise (in consistency with the number of independent identities). 
We finally come to the results presented in this paper. We will show that (\ref{S2id}) is equivalent to two identities, 
\be
\langle\textrm{out}\vert [Q_{r X},S] \vert\textrm{in}\rangle\ =\ 0, \label{sswardid}
\ee
\be
\langle\textrm{out}\vert [\tilde{Q}_{r X},S] \vert\textrm{in}\rangle\ =\ 0, \label{magsswardid}
\ee
associated to vector fields $\xi^a \sim r X^A\partial_A$. The charges are now parametrized by \emph{divergence-free} sphere vector fields $X^A$ and so \emph{each equation corresponds to one identity} (per point on the sphere). We will show that $Q_{r X}$ can be computed by phase space methods in the same way as done for $Q_V$ and $Q_f$. We currently lack a first principles derivation of $\tilde{Q}_{r X}$. From the structure of the leading and subleading cases in gravity and electromagnetism  we expect  that $\tilde{Q}_{r X}$ is the magnetic version of $Q_{r X}$. We will later comment further on this point, whose final clarification is left for future investigations.



We motivate our search  for  the new symmetry by looking at how soft theorem $\rightarrow$ Ward identities is accomplished in the known cases. To simplify computations  we  restrict attention to the case where the external particles are massless scalars. Let us for concreteness look at the leading soft theorem (\ref{S0id}). Using the relation between the graviton Fock operator and the Fourier transform of the   radiative free data \cite{aa}:
\be
 a_-(\w,\qh) = \sqrt{\gamma} \, 2 \pi i \, C^{zz}(\w,\qh),
 \ee
($\sqrt{\gamma}$ is the area element on the sphere of soft graviton directions $\qh$ parametrized by stereographic coordinates $(z,\zb)$) one can rewrite Eq. (\ref{S0id}) (for an outgoing negative helicity soft graviton) as:
\be
\sqrt{\gamma} 2 \pi i \lim_{\w \to 0} \w \langle\textrm{out}\vert C^{zz}(\w,\qh) S \vert\textrm{in}\rangle\ = S^{(0) -}  \langle\textrm{out} \vert S \vert\textrm{in}\rangle\ , \label{leadingst}
\ee
where $S^{(0) -} = \w \sum_{i=1}^{n}\frac{k_{i}^{\mu}\epsilon^{-}_{\mu\nu}k_{i}^{\nu}}{k_{i}\cdot q}$ is a function of the soft graviton direction $\qh$ and the external momenta $k_i$.  
On the other hand the `soft' (linear in $C^{zz}$) part of the supertranslation charge can be written as  \cite{stromST}:
\be
\Qs_f = i \lim_{\w \to 0} \; \w \int d^2 z \sqrt{\gamma} f D^2_z C^{zz}(\w,z,\zb) + c.c.
\ee
This motivates one to perform the operation $\int d^2 z f D^2_z$ on both sides of (\ref{leadingst}). The  identity
\be
\frac{1}{2 \pi}D_{z}^{2}S^{(0) -} = - \sum_{i=1}^n E_i \delta^{(2)}(z,z_{i}),
\ee
($z_i$ parametrizes the direction of the $i$-th external particle with energy $E_i$) allows one to write the right hand side  term as a local function of the external particle momenta and subsequently identify it with the action of the  `hard'  (quadratic) part of the supertranslation  charge $\Qh_f$. Thus, by smearing  both sides of the soft theorem (\ref{leadingst}) with $f D^2_{z}$ one arrives at (\ref{stwardid}) with $Q_f= \Qs_f + \Qh_f$.

Similar strategy applies to the subleading case where the appropriate smearing is $\int d^2 z V^z D^3_z$.
Whence the way to deduce the asymptotic charge from 
the soft theorem hinges on smearing both sides of the soft theorems with appropriate tensors.  We use the same logic to find the asymptotic charges 
from the sub-subleading soft theorem. We will then show that these charges are associated to certain symmetries of (perturbative) gravity.  

In the notation of Eq. (\ref{leadingst}) the sub-subleading relation  (\ref{S2id}) for a negative helicity soft graviton  takes the form
\be
\sqrt{\gamma} 2 \pi i \lim_{\w \to 0} \w^{-1} \langle\textrm{out}\vert C^{zz}(\w,\qh) S \vert\textrm{in}\rangle |_{\rm fin} = S^{(2) -}  \langle\textrm{out} \vert S \vert\textrm{in}\rangle\ , \label{subsubleadingst}
\ee
where $S^{(2) -} = \w^{-1} \sum_{i=1}^n (2\,  k_i \cdot q)^{-1}(\epsilon^\mu_- q^\nu J^i_{\mu \nu})^2$ is a function of $\qh$ and a differential operator on the external particles. Looking at the smearing employed in the leading and subleading cases, it is natural to attempt a smearing of the form $\int d^2 z Y^{zz}D^4_z$. One then finds an identity 
\be
\frac{1}{2 \pi} D_z^4 S^{(2) -} = -3 \sum_{i=1}^{n}  E_i^{-1}  \delta^{(2)}(z,z_{i}) \partial_{z_i}^2 + \ldots
\ee
in which all terms are proportional to (derivatives) of delta functions. Hence upon  smearing with $Y^{zz}D_{z}^{4}$ the right hand side of (\ref{subsubleadingst}) becomes a differential operator that is local in the external momenta. Furthermore, each term may be realized as the action of a hard charge $\Qh_Y$. Thus, just as in the case of the previous soft theorems  by smearing both sides of the sub-subleading theorem with $\int d^2 z Y^{zz}D_{z}^{4}$ we arrive at a relation of the form 
\be
\langle\textrm{out}\vert [\Q_{Y},S] \vert\textrm{in}\rangle\ =\ 0 \label{QYwardid}
\ee
where $\Q_Y = \Qcs_Y+\Qch_Y$ with \cite{long}:
\be\label{chargefromSSs}
\Qcs_{Y} = \int_{-\infty}^{\infty} du \int_{-\infty}^u du' \int d^2 z \sqrt{\gamma} \; Y^{zz} D_z^4 C^{zz}(u',\qh)  + c.c. \ee
\begin{multline}\label{chargefromSSh}
\Qch_Y = -  \int du \int d^2 z \sqrt{\gamma} \,\big( 3 Y^{zz} \partial_z \phi  \partial_z \phi  -   D^2_z Y^{zz} \phi^2 \\
 +2 u \, D_z Y^{zz} \partial_z \phi \, \partial_u \phi + \frac{u^2}{2}D^2_z Y^{zz} (\partial_u \phi)^2 \big) +c.c.
\end{multline}
The double integral in (\ref{chargefromSSs}) comes from the $\w^{-1}$ factor in  (\ref{subsubleadingst}). The field  $\phi$ in (\ref{chargefromSSh}) is the radiative data of the external massless particles.\footnote{Throughout the paper we assume $C_{AB}=O(u^{-2-\epsilon})$ and $\phi=O(u^{-1/2-\epsilon})$ at $u \to \pm \infty$ to ensure convergence of $u$ integrals.} As in the leading and subleading cases,  one can also go in the reverse direction  by an appropriate choice of $Y^{zz}$  and recover  (\ref{S2id}) from (\ref{QYwardid}). Note that we have only explicitly shown negative helicity contributions. The positive helicity terms appear in the complex conjugated (c.c.) piece. 

Our goal now is to show that such charges are associated to large spacetime diffeomorphisms. At first this may seem impossible as the charges are parametrized by $Y^{zz}$  or equivalently by  symmetric, trace-free sphere tensors $Y^{AB}$. However every such tensor can be written as  (symmetric, trace free part of) $D^A X^B$ for some sphere vector field $X^A$. 
We will show that for \emph{divergence-free} $X^A$ the charge $\Q_{Y^{AB}= D^A X^B}$ is associated  to a spacetime vector field with a leading $O(r)$ component  $\xi^a \sim r X^A \partial_A$. This however captures only `half' of the the $\Q_{Y}$ charges, the other half being labelled by  $Y^{AB}=\epsilon^{B}_{\;C} D^A X'^C$ with $X'^A$ divergence-free. This is the charge alluded to in Eq. (\ref{magsswardid}), namely
\be
\tilde{Q}_{r X'} :=\Q_{\epsilon^{B}_{\;C} D^A X'^C} .\label{defQtX}
\ee 
In short, using the splitting $Y^{AB}= D^A X^B + \epsilon^{B}_{\;C} D^A X'^C$ (with $X^A,X'^A$ divergence-free), the charges $\Q_Y$ are reinterpreted as a pair of charges $Q_{r X}$ and $\tilde{Q}_{r X'}$.\footnote{The situation is analogous to the subleading case in QED where the charges are parametrized by vector fields $Y^A$ \cite{stromlow}. For  $Y^A= D^A \mu$ the charge is associated to $O(r)$ large gauge transformations with leading piece $r \mu$. The magnetic dual of such charge is associated to  $Y^A= \epsilon^{AB} D_B \mu$ \cite{maxwell}.}

 
As generalised BMS symmetries are known to be equivalent to leading and subleading soft graviton theorems we know that we 
need a genuine extension of this group. Looking for such an extension is subtle in Bondi gauge as   generalised BMS 
appear 
to exhaust 
all such symmetries as far as smooth diffeomorphisms are concerned \cite{cl1}. Whence we look for such an extension in de Donder gauge. 
That is, we  look for vector fields on flat spacetime which satisfy the wave equation
\be\label{waveeqn}
\square\xi^{a}\ =\ 0.
\ee
The computation of asymptotic charges associated to symmetries in de Donder gauge also brings a nice structural coherence to the entire program. As the soft theorems are usually formulated in de Donder gauge as opposed to Bondi gauge, our analysis has a nice corollary which shows that the ``Ward identities $\equiv$ soft theorem" can be formulated in de Donder gauge for all  generators of the generalized BMS group \cite{long}.
In the present case, taking a cue from the large gauge transformations in QED which give rise to the subleading theorem \cite{maxwell} we look for large diffeomorphism generators in de Donder gauge such that the $O(r^{0})$ component of $\xi^{A}$ is linear in $u$.\\

It turns out that a self-consistent asymptotic solution compatible with the prescribed boundary behaviour 
is given by
\be \label{solxi}
\begin{array}{lll}
\xi^{A}\ =\ r X^{A}\ +\ \frac{u}{4}(\Delta +5) X^A\ +\ O(r^{-1})\\ 
\xi^{u}\ = O(r^{-1})  , \quad \xi^{r}\ =O(r^{-1}),
\end{array}
\ee
with $X^A$ a divergence-free, $u$-independent sphere vector field that plays the role of `independent data' in terms of which the remaining components are determined.   The form of the solution (\ref{solxi}) ensure the asymptotic charges satisfy certain regularity conditions detailed below.  The vector field  shares with generalized BMS generators  the property of being asymptotically divergence free, $\nabla_a \xi^a \to 0$ \cite{cl1}.

It is important to note  that at this stage we do not understand in what sense these large gauge transformations are symmetries of asymptotically flat spacetimes. Due to their diverging behavior at infinity, they naively do not seem to preserve asymptotic flatness \cite{long}. However as the Ward identities associated to their charges capture the sub-subleading soft theorem, we  believe  there should be a characterization of these large gauge transformations as symmetries of the theory. We leave this important question for future investigation.

We now proceed  to compute the  associated charges and show that they precisely yields the 
charge obtained from the sub-subleading theorem. The computation of charges is best done via 
covariant phase space techniques \cite{abr}.  Instead of considering pure gravity (for which the sub-subleading theorem is originally derived) we consider gravity coupled to massless scalar field  as it simplifies the calculations. 

In the context of tree-level amplitudes we are interested, it suffices to
 consider the phase space of linearized gravity coupled to the massless scalar field. Given a symmetry generator $\xi^a$, its associated charge has two contributions. One contribution  comes from the matter phase space and is given by 
\be \label{mattercharge}
Q^{\textrm{matter}}[\xi]\ =\ - \lim_{t\rightarrow\infty}\int_{\Sigma_t} d^3V  \, T^{t}_{\; b}\xi^{b},
\ee
where $\Sigma_t$ is a $t=$constant hypersurface approaching null infinity and $T^{a}_{\; b}$ 
the stress tensor of the scalar field. 
The other contribution comes from the gravitational phase space and is given by
\be
\delta Q^{\textrm{grav}}[\xi] \ =\lim_{t \to \infty} \delta_{\xi}\theta_t (\delta)\ -\ \delta\theta_t(\delta_{\xi})
\ee
where 
\be
\theta_t(\delta)\ =\ \frac{1}{2} \int_{\Sigma_t} d^3V \, \Gamma^{t}_{bc}\delta h^{bc}
\ee
is the linearized gravitational symplectic potential. Here $h_{ab}$ is the metric perturbation and 
$\Gamma^{a}_{bc}$ 
the linearized Christoffel connection.

As shown in \cite{long}, the computation of such a charge requires determining the linearized metric which is sourced by the matter stress tensor. As we are working in de Donder gauge, we need to analyze solutions to linearized Einstein's equations:
\begin{equation}
\begin{array}{lll}
\square \overline{h}_{ab}\ =\ -2T_{ab}
\end{array}
\end{equation}
where $\overline{h}_{ab}\ :=\ h_{ab}\ -\frac{1}{2}\eta^{mn}h_{mn}h_{ab}$.\\ 
A solution to this equation can be written as $\overline{h}_{ab}\ =\ \overline{h_{1}}_{ab} \ +\ \overline{h_{2}}_{ab}$ where $h_{1},\ h_{2}$ satisfy,
\be
\begin{array}{lll}
\square\overline{h_{1}}_{ab}\ =\ 0\\
\square\overline{h_{2}}_{ab}\ =\ -2T_{ab}.
\end{array}
\ee
Here $h_{1}$ is the linearized metric which is determined by the radiative data $C_{AB}$ at null infinity and $h_{2}$ is the linearized metric which is sourced by the matter and is independent of the radiative gravitational data.\\
The gravitational contribution to the charge is hence given by,
\be
\begin{array}{lll}
Q^{\textrm{grav}}[\xi] = Q_{\textrm{soft}}^{\textrm{grav}}[\xi] + Q_{\textrm{hard}}^{\textrm{grav}}[\xi]
\end{array}
\ee
where
\ba
Q_{\textrm{soft}}^{\textrm{grav}}[\xi] = \frac{1}{2}\lim_{t\rightarrow\infty}\int d^3 V \left(\Gamma^{t}_{ab}[h_{1}]\delta_{\xi}h^{ab} - \delta_{\xi}\Gamma^{t}_{ab}h_{1}^{ab}\right)\\
Q_{\textrm{hard}}^{\textrm{grav}}[\xi] = \frac{1}{2}\lim_{t\rightarrow\infty}\int d^3 V \left(\Gamma^{t}_{ab}[h_{2}]\delta_{\xi}h^{ab} - \delta_{\xi}\Gamma^{t}_{ab}h_{2}^{ab}\right)
\ea
The `soft' piece is linear in the radiative gravitational mode $C_{AB}$. 
The `hard' piece is linear in the matter  stress tensor and  
adds to the contribution coming from the matter phase space (\ref{mattercharge}).

Collecting all terms one finds the  resulting charge is divergent. However the nature of the divergent terms points to a natural prescription for obtaining the finite charge. More in detail, one finds
\be
Q[\xi]  = \lim_{t \rightarrow\infty} \left(  t \, Q^{(1)}[\xi] + Q^{(0)}[\xi]  \right). \label{Q10}
\ee
It turns out that $ Q^{(1)}[\xi]$ is the  (hard) charge associated to generalized BMS sphere vector fields.  Thus, extracting the finite piece in (\ref{Q10}) amounts to discarding the contributions from the subleading soft gravitons   (see \cite{maxwell} for similar prescription in QED).   It is at this stage that the form (\ref{solxi}) of $\xi^a$  is crucial: Other vector fields satisfying (\ref{waveeqn}) yield divergent contributions that cannot be associated to  generalized BMS charges. 

The total, finite charge associated to  $\xi^a$ is finally given by $Q_\xi:=Q^{(0)}[\xi]= Q^{\textrm{hard}}_{\xi} + Q^{\textrm{soft}}_{\xi}$ with \cite{long}:
\be \label{Qsfinal}
Q^{\textrm{soft}}_{\xi}\ = \frac{1}{16} \int d u \, d^2 V \,  s^{AB}(\xh) \int_{-\infty}^u C_{AB}(u',\xh) du' ,
\ee
\be
Q^{\textrm{hard}}_{\xi}=  \frac{1}{4} \int d u \, d^2 V \, \big( 3 D^A X^B T^{(-2)}_{AB} +  u (\Delta X^A +X^A) T^{(-2)}_{u A} \big),\label{Qhfinal}
\ee
where:
\be
s^{AB}=  \Delta^2 D^A X^B - 6 \Delta D^A X^B +8 D^A X^B ,
\ee 
 and  $T^{(-2)}_{AB}= (\partial_A \phi \, \partial_B \phi)^{\tf}$ , $T^{(-2)}_{u A}= \partial_u \phi \,  \partial_A \phi$ are leading terms of the stress tensor.  Upon the identification $Y^{AB}=-\frac{1}{4}D^A X^B$ and using $D_A X^A=0$ one finds that (\ref{Qsfinal}) and (\ref{Qhfinal}) exactly match the respective charges (\ref{chargefromSSs}) and (\ref{chargefromSSh}) that were obtained from the  sub-subleading theorem.

As the large gauge transformations are parametrized by divergence free-vector fields on the sphere it corresponds to one  factorization theorem for each direction of soft graviton as opposed to the two factorization theorems given in Eq.(\ref{S2id}). Whence we are missing ``half'' of the Ward identities which would correspond to the remaining half of the sub-subleading theorem. It is here that we take a motivation from \cite{maxwell} where it is shown that a single large gauge transformation gives rise to both sub-leading relations (for two photon helicities) in massless QED. 

This is due to the fact that the magnetic and electric charge for a large gauge transformation are unequal and their Ward identities are equivalent to the Low's subleading theorem. Whence a  couple of questions naturally arise: (i) Is the charge $Q_\xi$ we have computed analogous to the electric charge in the case of QED? and if it is (ii) What is the corresponding magnetic charge? There are strong reasons to believe that the answer to the first question is in affirmative due to a reinterpretation of generalized BMS charges  presented in \cite{long}. As shown there, these charges can be obtained from the ``electric'' part of the Weyl tensor. Electric and magnetic part of the Weyl tensor whose leading piece contains information about radiative mode can be defined as\footnote{This is not the standard definition of ${\cal E}$ and ${\cal B}$  but it  contains complete information of the Weyl tensor and at null infinity have trivial projection in the outgoing null direction.}
\be
{\cal E}^{a}_{b}\ :=\ r \,C^{at}_{\phantom{at} br} ,\quad \quad {\cal B}^{a}_{b}\ :=\ r *C^{at}_{\phantom{at} br}.
\ee
Using these tensors, the generalized BMS charges can be obtained as
\begin{equation}
Q_{\cal E}[\xi]\ =\ \lim_{t\rightarrow\infty}\int_{\Sigma_{t}} d^3 V\, \partial_{a}({\cal E}^{a}_{b}\xi^{b}). \label{QE}
\end{equation}
The `magnetic' dual charges are then defined by replacing ${\cal E}^a_b$ with ${\cal B}^a_b$ in (\ref{QE}). Thus for
a supertranslation vector field  $\xi^{a}_f \sim f \partial_u$, $Q_{\cal E}[\xi_f]$ reproduces the usual supertranslation charge. 
The corresponding magnetic charge turns out to be:\footnote{This charge was first obtained in \cite{nut4} by conformal methods.}
\ba
 Q_{{\cal B}}[\xi_f]\ =  \int du \int d^2 V f \epsilon^{AB} D_A D^M \partial_u C_{MB} .
\ea
The charge is linear in the graviton and putting it equal to zero precisely gives the so-called CK condition  used in \cite{stromST}. For the remaining generators of generalized BMS, $\xi^a_V \sim V^A \partial_A$,  it can also be shown that the electric charge matches the $\textrm{Diff}(S^{2})$ charge computed in \cite{cl1} (in this case magnetic charges turn out to coincide with the electric charges).  Whence we expect that the charge associated to $\xi^a \sim r X^A \partial_A$ computed above can be obtained as an electric charge  whose magnetic counterpart provides the `remaining' information of the $\Q_Y$ charges, Eq. (\ref{defQtX}). However a detailed proof of this statement remains outside the scope of this work.

We thus believe to have provided enough evidence to prove that the extensions of the BMS algebra previously considered in the literature are not the end of the story. The sub-subleading theorem of tree level quantum gravity amplitude implies the existence of a further extension of such algebras to a potentially larger symmetry. However in what sense 
this extension is
a symmetry of  asymptotically flat spacetimes and hence whether sub-subleading soft gravitons can also be understood as Goldstone modes of a spontaneously broken symmetry remains to be seen. \\

We would like to thank Freddy Cachazo for his suggestion and encouragement to look for a symmetry interpretation of the sub-subleading soft graviton theorem. We are grateful to Abhay  Ashtekar for many discussions on asymptotic symmetries in gravity and on the importance of magnetic charges. MC would like to thank Philipp Hoehn for an invitation to Perimeter Institute and to FC for stimulating discussions during the visit. MC also thanks Rodrigo Eyheralde, Rodolfo Gambini and Michael Reisenberger for useful discussions. AL is grateful to Arnab Priya Saha for helpful discussions. 
MC is supported by Anii and Pedeciba. AL is supported by Ramanujan Fellowship of the Department of Science and Technology.

\end{document}